\documentstyle[12pt,epsfig]{article}

\newcommand{\beq}{\begin{equation}}
\newcommand{\eeq}{\end{equation}}
\newcommand{\bea}{\begin{eqnarray}}
\newcommand{\eea}{\end{eqnarray}}

\def\matz   {|\overline{\cal{M}}_3|^2}
\def\wh{\widehat}

\def\lapprox{\lower .7ex\hbox{$\;\stackrel{\textstyle <}{\sim}\;$}}

\def\as     {\alpha_s}

\def\shift  {\rule[-3mm]{0mm}{8mm}}

\def\gapprox{\lower .7ex\hbox{$\;\stackrel{\textstyle >}{\sim}\;$}}

\begin{document}
\titlepage
\begin{flushright}
{DTP/97/66}\\
{hep-ph/9707373}\\
{July 1997}\\
\end{flushright}

\begin{center}

\vspace*{2cm}

{\Large {\bf Radiation Zeros at HERA --- \\[2mm]
More About Nothing}} \\

\vspace*{1.5cm}
M.~Heyssler$^{a}$ and W.J.~Stirling$^{a,b}$ \\

\vspace*{0.5cm}
$^a \; $ {\it Department of Physics, University of Durham,
Durham, DH1 3LE }\\

$^b \; $ {\it Department of Mathematical Sciences, 
University of Durham, Durham, DH1 3LE }

\end{center}

\vspace*{4cm}

\begin{abstract}
The  process  $eq  \to eq  +\gamma$  exhibits  radiation  zeros,  i.e.
configurations of the final--state  particles for which the scattering
amplitude  vanishes.  We study these zeros for both  $e^+u$ and $e^+d$
scattering.  The latter exhibits a type of zero which to our knowledge
has not previously been  identified.  The  observability  of radiation
zeros at HERA is discussed.
\end{abstract}

\newpage

\section{Introduction}

In certain high--energy scattering processes involving the emission of
one or more photons, the scattering  amplitude vanishes for particular
configurations of the final--state particles.  Such configurations are
known  as  {\it  radiation  zeros}.  In the  context  of  high--energy
scattering, they were first discussed by Mikaelian,  Sahdev and Samuel
\cite{Mik79}.  To  measure  the  magnetic  moment  $\mu_W$  of the $W$
boson, they  proposed and studied the $W$ boson  production  processes
$u\bar{d}\rightarrow  W^+\gamma$ and $d\bar{u}\rightarrow  W^-\gamma$.
They found that the matrix element  vanishes at a particular  value of
the c.m.s. frame scattering angle $\cos\wh{\theta}_W = [e_{d(\bar{d})}
-   e_{\bar{u}(u)}]/[e_{d(\bar{d})}   +   e_{\bar{u}(u)}]   =   -1/3$,
independent  of the  photon  energy.  A similar  effect is seen in the
$W\rightarrow     q\bar{q}\gamma$    decay    process    \cite{Gro81}.
Experimentally,  these radiation zeros have been observed  recently by
the  CDF  collaboration  \cite{CDF97}  at  the  Tevatron  $p  \bar  p$
collider.  There has also been renewed theoretical interest, including
studies on double  photon  emission  processes  $p\bar{p}  \rightarrow
W^{\pm}\gamma\gamma \rightarrow \ell^\pm\nu\gamma\gamma$  \cite{Bau97}
and  the  uniqueness  of  radiation   zeros  to  the  Standard   Model
\cite{Abr97}.  The energy  dependence  \cite{Rei86} of radiation zeros
in $pp \rightarrow  \gamma + X$ \cite{Rei89} and using radiation zeros
to probe the  colour--charge  of  partons  \cite{Hag84}  has also been
studied.  A review of recent  developments in the subject can be found
in Ref.~\cite{Bro95}.

A first  understanding of the phenomenon of radiation  amplitude zeros
was  achieved  in  the  pioneering  work  of   Ref.~\cite{Bro82}.  The
vanishing of the  scattering  amplitude can be  understood  as arising
from  complete  destructive  interference  of the classical  radiation
patterns of the incoming and outgoing  charged  lines in  relativistic
$n$--particle  collisions.  Taking the single  emission of a photon as
the paradigm  process, it was shown that the amplitudes  can 
vanish if the
other  particles  participating  in the process  have the same sign of
charge  $e_i$.\footnote{In  fact  in  general  this  is only  true  at
tree  level, see Ref.~\cite{Bro82}.}

Same--sign  charge scattering occurs naturally in high--energy  hadron
collisions in subprocesses  such as $u\bar d \to W^+ \gamma$.  However
similar phenomena can be expected in lepton--hadron collisions, and in
particular at HERA in processes such as $eq \rightarrow eq+\gamma$ for
$eq = e^+u$ or $e^-d$.  Studies of radiation zeros for these processes
at HERA were first performed by Bilchak \cite{Bil85} and more recently
by Doncheski and Halzen \cite{Don91}.

In a recent paper  \cite{Hey97} we studied the  distributions  of soft
gluon and photon  radiation  in $eq \to eq$  scattering  at HERA.  The
motivation  was  to  demonstrate  that  the  radiation   patterns  are
different depending on whether the scattering takes place via standard
$t$--channel $\gamma^*,Z^*$ exchange  or via the  production  of a new
heavy, charged, colour--triplet `leptoquark' (LQ) resonance in the $s$
channel.  Leptoquark   production   is   one   of   several   possible
explanations  for the apparent  excess of high--$Q^2$  deep  inelastic
scattering events at HERA \cite{H1.97,ZEUS97}.

A by--product of this study was the  identification of radiation zeros
in both the  Standard  Model and  leptoquark  $e^+q \to e^+q + \gamma$
($q=u,d$)   scattering   amplitudes.  For  a   long--lived   resonance
($\Gamma_{\rm  LQ}\rightarrow  0  $)  we  found  radiation  zeros  for
scattering of particles with the same sign (i.e.  $e^+u$ scattering in
our case) and zeros outside the physical region for $e^+d$ scattering,
consistent  with  the  results  derived  in  Refs.~\cite{Bil85,Don91}.
However we also found that for a short--lived resonance  ($\Gamma_{\rm
LQ}\rightarrow  \infty $) {\it and for the Standard  Model} there were
radiation zeros also for $e^+d$ scattering within the physical region.
Both types of Standard Model $e^+q \to  e^+q+\gamma$  radiation  zeros
will be the focus of the present study.

From an  experimental  point of view the  detection  of photons in the
final state is highly  non--trivial.  The rates are small  (suppressed
by ${\cal  O}(\alpha)$  compared to the total cross  section)  and the
photons  must be  well--separated  from  the beam and from  the  other
final--state  particles, and contained within the detector.  The basic
question is whether the radiation  zeros of the  scattering  amplitude
correspond  to  `detectable'  photons  at HERA.  In this study we will
present  results for typical values of the DIS variables $y$ and $Q^2$
which  correspond to  observable  quark jets and scattered  positrons.
For these values we will  investigate  the  location of the  radiation
zeros for photons with an energy greater than 5~GeV.

The paper is  organised  as follows.  We first  consider  soft--photon
emission  and  derive  analytic  solutions  for  the  location  of the
radiation  zeros  in the  $eq$  c.m.s.  frame.  We then  show  how the
transition from soft-- to hard--photon emission shifts the position of
the  zeros.  Finally  we move to the HERA lab frame to see  where  the
zeros   occur   in  the   detector.  We   also   compare   our   exact
matrix--element  results  with an  approximate  calculation  in  which
photon  emission is included  in the  collinear  approximation,  which
could  correspond for example to a  parton--shower  implementation  of
such  emission.  This  model has no  radiation  zeros and  serves as a
benchmark for the amplitude suppression in the exact result.

\section{Radiation zeros in $eq\to eq\gamma$ scattering}

In the following we shall study the reactions
\begin{eqnarray}  \label{eq:reaceu} 
e^+(p_1) \; u(p_2) &\rightarrow&
e^+(p_3)  \;  u(p_4)  +  \gamma(k),   
\\
e^+(p_1)  \; d(p_2)  &\rightarrow&  
e^+(p_3)  \;  d(p_4) +  \gamma(k).
\label{eq:reaced}  
\end{eqnarray}  
Other  scattering  combinations  ($e^+\bar  u, e^- u, \ldots$)  can be
obtained from these basic processes by readjusting the charge factors.
The  expression  for the matrix element  squared  (summed and averaged
over spins) may for example be obtained by crossing the expression for
$e^+e^-\rightarrow  \mu^+\mu^-+\gamma$ given in Ref.~\cite{Ber81}.  In
terms       of       the        four--momenta        defined        in
Eqs.~(\ref{eq:reaceu},\ref{eq:reaced}) the matrix element for massless
quarks and leptons is
\beq \label{eq:fullmat}
\matz(e^+q\rightarrow e^+q+\gamma) = e^6e_q^2
\frac{(p_1\cdot p_2)^2+(p_3\cdot p_4)^2+(p_1\cdot p_4)^2+
(p_2\cdot p_3)^2}{(p_1\cdot p_3)(p_2\cdot p_4)}
{\cal{F}}^\gamma_{\rm SM},
\eeq
with
\beq \label{eq:fsm}
\frac{1}{2}{\cal{F}}^\gamma_{\rm SM} = e_q^2 [24] - 
e_q \left\{ [12] + [34] - [14] - [23] \right\} + [13].
\eeq
We have used here the following  short--hand  notation for the eikonal
factors:
\beq \label{eq:eikonal}
[ij] = \frac{p_i\cdot p_j}{(p_i \cdot k)(p_j \cdot k)}.
\eeq
The expression in Eq.~(\ref{eq:fsm})  --- the {\em antenna pattern} of
the process --- contains collinear $(\vec{{\rm \bf k}} \cdot \vec{{\rm
\bf p}}_i \rightarrow 0)$ as well as infrared  ($\omega_\gamma  \equiv
E_k \rightarrow  0$)  singularities.  It is this factor which vanishes
for certain configurations of the momenta.  Note that we only take the
neutral  current  $\gamma^*$--exchange  into  account  as the  antenna
pattern  in   Eq.~(\ref{eq:fsm})   is  independent  of  the  exchanged
particles  as  long  as they  do not  themselves  emit  photons.  This
approximation  will  influence the cross section rate slightly at high
$Q^2$, but will not affect the position of the radiation zeros.

\subsection{Type 1 radiation zeros}

\begin{figure}[t]
\begin{center}
\mbox{\epsfig{figure=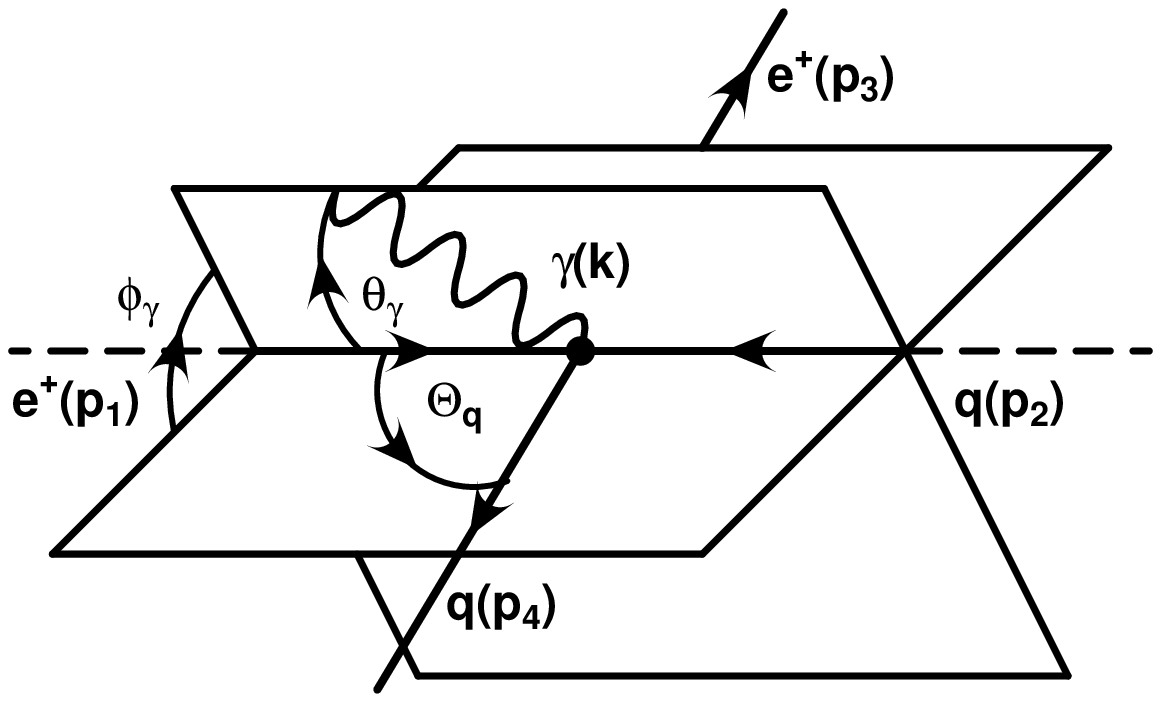,width=11.5cm}}
\caption[]{Parametrisation  of the  kinematics  for  $e^+(p_1)  q(p_2)
\rightarrow  e^+(p_3)  q(p_4) +  \gamma(k)$  scattering  in the $e^+q$
c.m.s.  frame.  The   orientation  of  the  photon   relative  to  the
scattering  plane is denoted  by  $\theta_\gamma$  and  $\phi_\gamma$.
}
\label{fig:kinemat}
\end{center}
\end{figure}

\begin{figure}[t]
\begin{center}
\mbox{\epsfig{figure=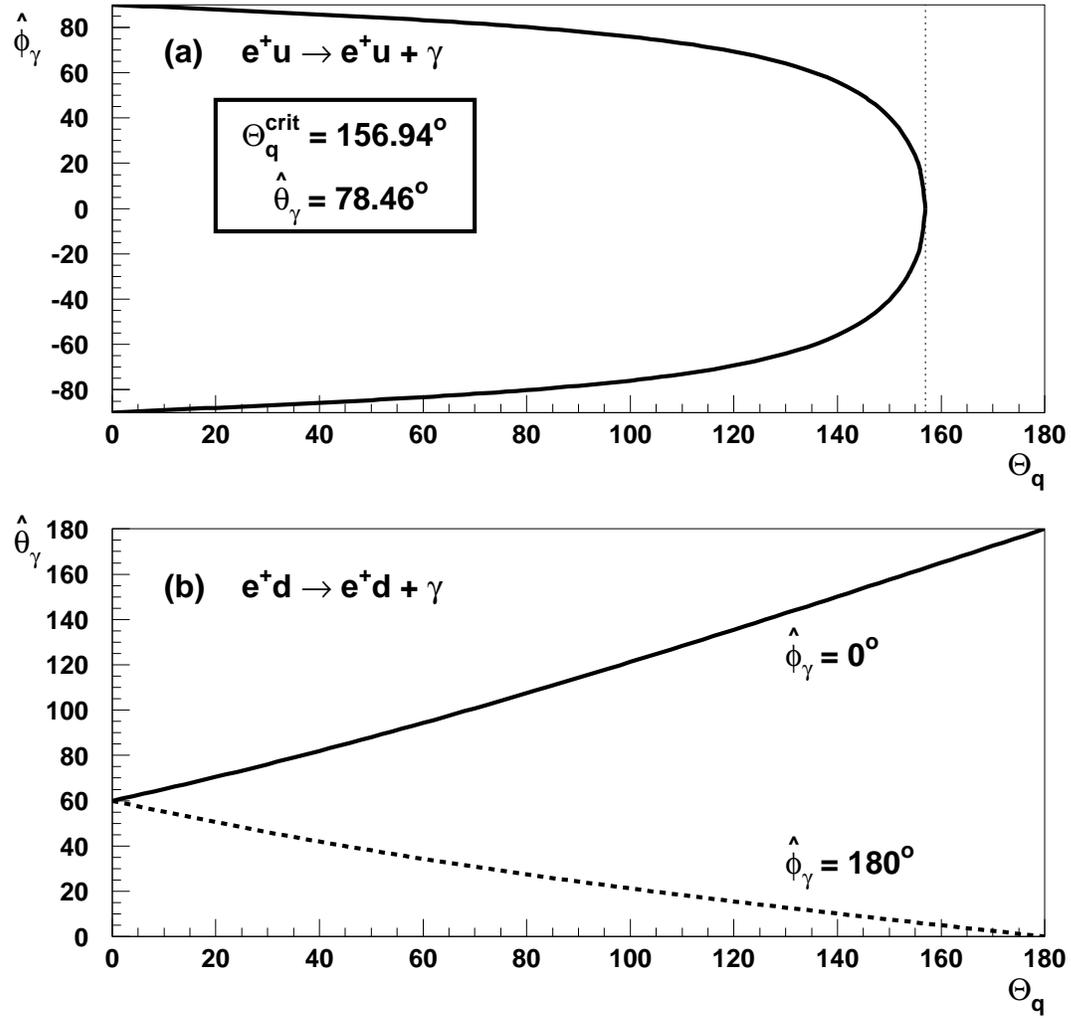,width=15.5cm}}
\caption[]{The  position of the  radiation  zeros as a function of the
quark  scattering  angle $\Theta_q$ for  soft--photon  emission in (a)
$e^+u \rightarrow e^+u+\gamma$ and (b) $e^+d \rightarrow  e^+d+\gamma$
in the  $(\phi_\gamma,\theta_\gamma)$  c.m.s.  phase space of the soft
photon.}
\label{fig:cmszeros}
\end{center}
\end{figure}

To see under what conditions  ${\cal{F}}^\gamma_{\rm SM}$ vanishes, we
first recall the `single--photon theorem' from Ref.~\cite{Bro82} which
states that the amplitude  vanishes when the  charge--weighted  scalar
products  $Q_i/(p_i\cdot k)$ are equal.  If we denote the common value
by $\lambda$, then
\beq \label{eq:eikonal2}
[ij] = (Q_i Q_j)^{-1} \lambda^2 \; p_i\cdot p_j
\eeq
and   it   is    straightforward    to   show   by   substitution   in
Eq.~(\ref{eq:fsm}) that this gives ${\cal{F}}^\gamma_{\rm SM}= 0$.  In
the present  context,  the  equality  of the  charge--weighted  scalar
products corresponds to
\beq \label{eq:conditions}
\frac{1}{p_1\cdot k} = 
\frac{e_q }{p_2\cdot k} = 
\frac{1}{p_3\cdot k} = 
\frac{e_q}{p_4\cdot k}.  
\eeq
We can obtain a simple analytic solution to these equations by  taking
the soft--photon limit in which $\omega_\gamma/E_i \rightarrow 0$.  In
this limit we have  simple  two--body  kinematics  for the quarks  and
leptons,  $p_1 + p_2 = p_3 + p_4$.  If we  work in the  $e^+q$  c.m.s.
frame, and define  $\theta_2,  \theta_4$  to be the angle  between the
photon and the incoming  and outgoing  quarks  respectively,  then the
equations (\ref{eq:conditions}) become
\beq \label{eq:conditions2}
\frac{1}{1+z_2} = 
\frac{e_q }{1-z_2} = 
\frac{1}{1+z_4} = 
\frac{e_q}{1-z_4},  
\eeq
where $z_i = \cos\theta_i$. Equivalently,
\beq \label{eq:solut}
z_2 = z_4 = \frac{1-e_q}{1+e_q}.
\eeq
A necessary  condition for such a solution to physically exist is $e_q
\geq 0$  ($\Rightarrow  | z_i | \leq 1$), i.e.  $e^+u$ or  $e^+\bar d$
scattering.  This reproduces the well--known  result for scattering of
particles  with the same sign of  electric  charge,  as  discussed  in
Refs.~\cite{Bro82}.  We call these {\bf Type 1}  radiation  zeros.  By
itself,  however,  the  condition  $e_q \geq 0$ is not  sufficient  to
guarantee a zero in the  scattering  amplitude.  The  equation  $z_2 =
z_4$  can  only  be  satisfied  for  certain   configurations  of  the
final--state   particles.  To  see  this,  we  introduce  an  explicit
representation of the c.m.s.  four--momenta:
\begin{eqnarray}
p_1^\mu &=& \frac{\sqrt{\hat{s}}}{2} \left( 1, 0, 0, -1 \right) ,
\\
p_2^\mu &=& \frac{\sqrt{\hat{s}}}{2} \left( 1, 0, 0, 1 \right) ,
\\
p_4^\mu &=& \frac{\sqrt{\hat{s}}}{2} \left(1, \sin\Theta_q, 0, 
\cos\Theta_q \right) ,
\\
p_3^\mu &=&  \frac{\sqrt{\hat{s}}}{2} \left(1, -\sin\Theta_q, 0, 
- \cos\Theta_q \right)  ,
\\
k^\mu &=& \omega_\gamma \left(1, \sin\theta_\gamma \cos\phi_\gamma, 
\sin\theta_\gamma\sin\phi_\gamma,
\cos\theta_\gamma \right).
\end{eqnarray}
These  variables  are  illustrated  in  Fig.~\ref{fig:kinemat}.  It is
straightforward to show that the conditions for ${\cal{F}}^\gamma_{\rm
SM}=0$ defined in Eq.~(\ref{eq:solut}) correspond to
\beq \label{eq:uzeros}
\cos\wh{\theta}_\gamma = \frac{1-e_q}{1+e_q},
\eeq
and 
\beq \label{eq:uzerosphi}
\wh{\phi}_\gamma = 
\pm \arccos\left( \frac{\tan(\Theta_q/2)}{\tan\wh{\theta}_\gamma} \right).
\eeq
Thus for $e_u = +2/3$ we find radiation  zeros at  $\wh{\theta}_\gamma
\simeq    78.46^{\circ}$    and   for    $e_{\bar{d}}   =   +1/3$   at
$\wh{\theta}_\gamma  =  60^{\circ}$.  We present the  positions of the
radiation  zeros  $(\wh{\phi_\gamma},\wh{\theta_\gamma})$  for process
(\ref{eq:reaceu})  ($e^+u$  scattering)  in  Fig.~\ref{fig:cmszeros}a.
Note   that   the    requirement   of   a   physical    solution   for
$\wh{\phi}_\gamma$  places  restrictions on $\Theta_q$.  There are two
radiation  zeros  in  the   $(\phi_\gamma,\theta_\gamma)$   plane  for
$\Theta_q  <  2\wh{\theta}_\gamma  \simeq  156.94^{\circ}$.  The cones
around the incoming  and outgoing  quarks  defined by $z_2, z_4 = 1/5$
have  two  lines  of  intersection  along  which  there is  completely
destructive   interference   of  the  radiation.  Note  also  that  at
$\Theta_q =  2\wh{\theta}_\gamma  = \Theta_q^{\rm crit}$ the radiation
zeros   degenerate   to  a   single   line   (i.e.  single   point  in
$(\phi_\gamma,\theta_\gamma)$  space) located in the scattering  plane
$(\wh{\phi}_\gamma=0^\circ)$.  There  are  no   radiation   zeros  for
$\Theta_q > 2\wh{\theta}_\gamma  \simeq 156.94^{\circ}$.  Finally, for
$\Theta_q=0^{\circ}$  there is an infinite  number of radiation  zeros
(`null  zone')  located on a cone  around  the beam line with  opening
angle $\wh{\theta}_\gamma$.

\subsection{Type 2 radiation zeros}

The processes (\ref{eq:reaceu},\ref{eq:reaced}) exhibit a second class
of  radiation  zeros, which we call {\bf Type 2}, which do not satisfy
the   `single--photon   theorem'.  These  zeros  are  located  in  the
scattering    plane    at     $\wh{\phi}_\gamma=     0^{\circ}$    and
$\wh{\phi}_\gamma=        180^{\circ}$.       The        corresponding
$\wh{\theta}_\gamma$ values may be calculated straightforwardly in the
soft--photon approximation as a function of the quark charge $e_q$ and
the quark scattering angle $\Theta_q$.  The result is
\beq \label{eq:dzeros}
\cos\wh{\theta}_\gamma = \frac{1}{2} 
\frac{\left( 1 - e_q^2 \right)
\left( 1 + \cos\Theta_q \right) \pm 
\sqrt{\Delta_\gamma
\left( e_q,\cos\Theta_q\right)}
}
{\left( 1 - e_q \right)^2},
\eeq
with
\beq 
\label{eq:dzerosdelta}
\Delta_\gamma
\left( e_q,\cos\Theta_q\right) =
\left[ \left( e_q^2-1\right)
\left( 1 + \cos\Theta_q \right) \right]^2 -
4\left(1-e_q\right)^2
\left(e_q^2\cos\Theta_q+2e_q+\cos\Theta_q\right).
\\ 
\eeq
The  condition  $\Delta_\gamma\left(  e_q,\cos\Theta_q\right)  \geq 0$
constrains  the range of $e_q$ for  which  physical  zeros  exist.  In
terms of the polar angle $\Theta_q$ we have
\beq \label{eq:eqrange1}
-\infty < e_q \leq \frac{\cos\Theta_q + 
3 -2\sqrt{2(1+\cos\Theta_q)}}
{1-\cos\Theta_q} \leq 1,
\eeq
or
\beq 
\label{eq:eqrange2} 1\leq 
\frac{\cos\Theta_q + 3 +2\sqrt{2(1+\cos\Theta_q)}}
{1-\cos\Theta_q}\leq e_q < +\infty,
\eeq
the latter being actually  redundant  since Standard Model quarks have
$|e_q|\leq +2/3$.  From  Eq.~(\ref{eq:eqrange1}) we obtain constraints
on the quark  scattering angle  $\Theta_q$ for particular  flavours of
quark.  There are radiation  zeros for all $e_q<0$ and for  positively
charged  quarks in a limited  range of  $\Theta_q$.  We summarise  the
results in Table~\ref{tab:thetaxrange}.
\begin{table}[htb]
\begin{center}
\begin{tabular}{|c|l|l|l|} \hline
\shift $e^+u$                                                & 
$e_q = +2/3$                                                 & 
$\cos\Theta_q \leq \pi - \arccos\left( \frac{23}{25}\right)$ & 
$\Theta_q \gapprox 157^{\circ}$                           \\ \hline
\shift $e^+\bar{d}$                                          & 
$e_q = +1/3$                                                 & 
$\cos\Theta_q \leq -\frac{1}{2}$                             & 
$\Theta_q \geq 120^{\circ}$                               \\ \hline
\shift $e^+d$                                                & 
$e_q = -1/3$                                                 & 
$\forall\; \cos\Theta_q$                                     & 
$\forall\;\Theta_q$                                       \\ \hline
\shift $e^+\bar{u}$                                          & 
$e_q = -2/3$                                                 & 
$\forall \; \cos\Theta_q$                                    & 
$\forall\;\Theta_q$                                       \\ \hline
\end{tabular}
\caption[]{{\em  Ranges of the quark scattering angle  $\Theta_q$, for
different  quark charges, for which  radiation zeros exist.  Note that
for $e_q<0$  there are always two  radiation  zeros in the  scattering
plane   for   $\wh{\phi}_\gamma=(0^{\circ},180^{\circ})$    with   the
$\wh{\theta}_\gamma$ value given by Eq.~(\ref{eq:dzeros}).}}
\label{tab:thetaxrange}
\end{center}
\end{table} 
Note that $e^+u$ scattering has both Type 1 and 2 zeros.  However, the
latter are  located  very close to the beam  direction,  making  their
observation  difficult in practice.  They also require very high $Q^2$
(back--scattered  quarks) and therefore  have a small event rate.  The
positions of the  Type 2 zeros  for   $e^+d$  scattering are shown  in
Fig.~\ref{fig:cmszeros}b   as  a  function  of   $\Theta_q$.  Finally,
Table~\ref{tab:cmsxzeros}  lists the numerical values of the radiation
zero  angles  $(\wh{\phi}_\gamma,   \wh{\theta}_\gamma)$  for  several
values of $\Theta_q$.
\begin{table}[htb]
\begin{center}
\begin{tabular}{|c||l|l||l|l|} \hline
\shift $\Theta_q$                        & 
\multicolumn{2}{c||}{$e^+d$ scattering}  & 
\multicolumn{2}{c|}{$e^+u$ scattering}\\ \hline
\shift $30^{\circ}$                      & 
$(0^{\circ},76.12^{\circ})$              & 
$(180^{\circ},46.12^{\circ})$            &
$(-86.86^{\circ},78.46^{\circ})$         & 
$(86.86^{\circ},78.46^{\circ})$       \\ \hline
\shift $45^{\circ}$                      & 
$(0^{\circ},84.98^{\circ})$              & 
$(180^{\circ},39.98^{\circ})$            &
$(-83.23^{\circ},78.46^{\circ})$         & 
$(83.23^{\circ},78.46^{\circ})$       \\ \hline
\shift $90^{\circ}$                      & 
$(0^{\circ},114.29^{\circ})$             & 
$(180^{\circ},24.29^{\circ})$            &
$(-78.22^{\circ},78.46^{\circ})$         & 
$(78.22^{\circ},78.46^{\circ})$       \\ \hline
\end{tabular}
\caption[]{{\em  Position of the radiation  zeros  $(\wh{\phi}_\gamma,
\wh{\theta}_\gamma)$  for  three  different  quark  scattering  angles
$\Theta_q$, in the soft--photon approximation.  }}
\label{tab:cmsxzeros}
\end{center}
\end{table} 

\subsection{Radiation zeros for arbitrary photon energies}

The   analytic   results   obtained   above   use   the   soft--photon
approximation.  However  radiation  zeros of both types exist for {\em
all} photon  energies and can be located using  numerical  techniques.
We  continue  to work in the  $e^+q$  c.m.s.  frame but now use  exact
$2\rightarrow   3$   kinematics.  Without   any   essential   loss  of
generality,  we can keep the  direction  $(\Theta_q)$  and the  energy
$(E^\prime_q)$  of the outgoing quark fixed and vary the direction and
energy  of  the  outgoing  photon,  constructing   simultaneously  the
four--momentum  of  the  outgoing  positron  to  conserve  energy  and
momentum.  The new  four--vectors  of the outgoing  quark,  lepton and
photon momenta are then
\begin{eqnarray} 
p_4^\mu    &=& 
E^\prime_q \left(1,\sin\Theta_q,0,\cos\Theta_q\right) ,
\label{eq:kincms2.p4}
\\
p_3^\mu    &=& p_1^\mu +p_2^\mu 
-p_4^\mu-k^\mu ,
\label{eq:kincms2.p3}
\\
k^\mu      &=& 
\omega_\gamma \left(1,\sin\theta_\gamma\cos\phi_\gamma,
\sin\theta_\gamma\sin\phi_\gamma,\cos\theta_\gamma\right) .
\label{eq:kincms2.k}
\end{eqnarray}

Once   again   we   obtain   a    vanishing    matrix    element    in
Eq.~(\ref{eq:fullmat})   if  the  antenna   pattern   ${\cal   F}_{\rm
SM}^\gamma$  of  Eq.~(\ref{eq:fsm})  is  zero.  For  Type 1  radiation
zeros, the  single--photon  theorem again leads to the  conditions  in
Eq.~(\ref{eq:conditions}).  The  equality  of  $p_1\cdot  k$ and  $p_2
\cdot  k$  leads  immediately  to   Eq.~(\ref{eq:uzeros}),   i.e.  the
radiation zeros are at fixed  $\wh{\theta}_\gamma$ {\em independent of
the photon  energy}.  However the azimuthal  angle  $\wh{\phi}_\gamma$
does vary with $\omega_\gamma$, since the supplementary condition $z_2
= z_4$ only  applies in the  $\omega_\gamma  \to 0$ limit.  For Type 2
zeros, it can be shown that the condition $\wh{\phi}_\gamma = 0^\circ,
180^\circ$  again  applies for  arbitrary  $\omega_\gamma$,  i.e.  the
zeros are always located in the scattering plane.

\begin{figure}[t]
\begin{center}
\mbox{\epsfig{figure=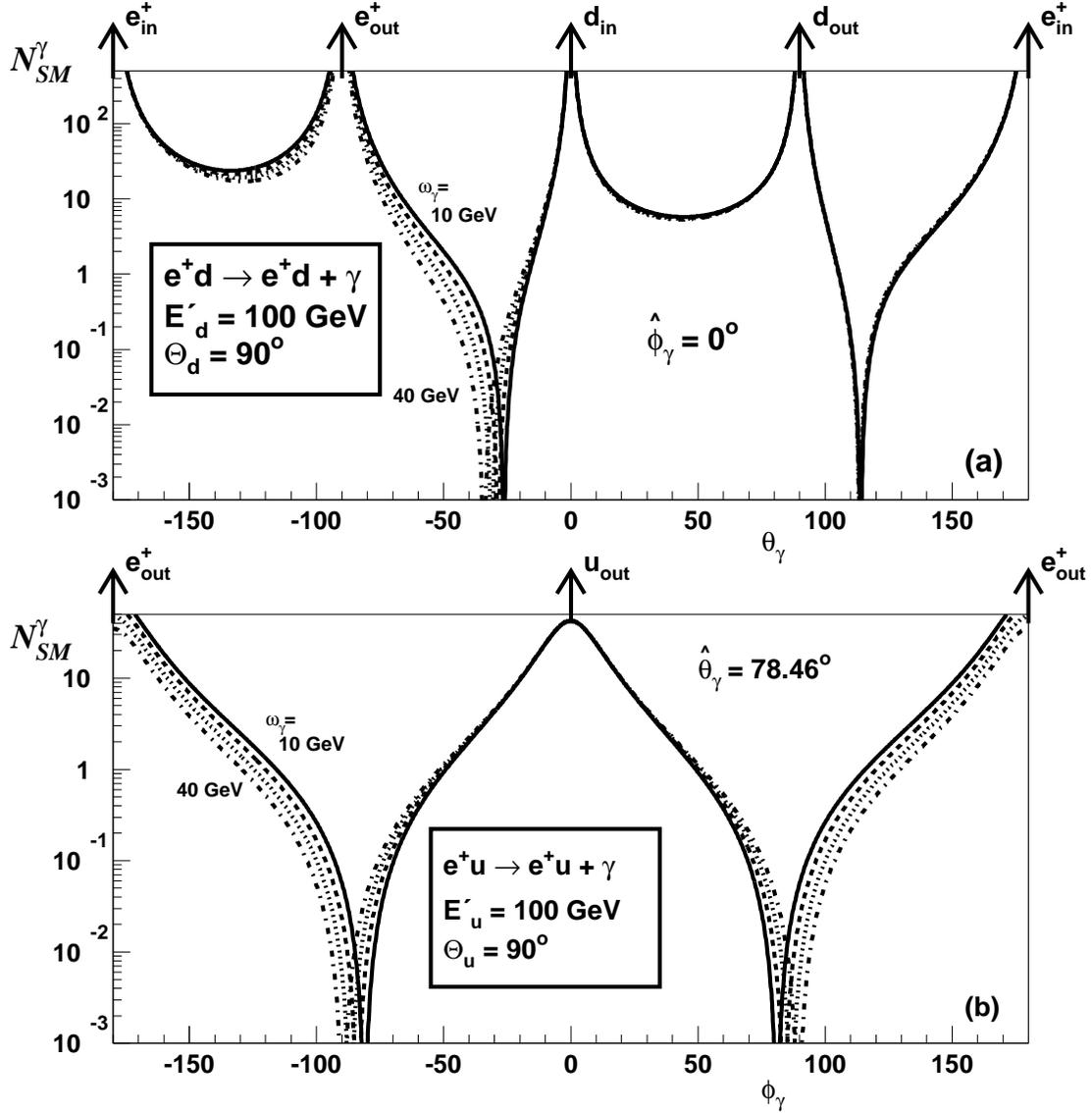,width=15.5cm}}
\caption[]{The dimensionless antenna pattern ${\cal N}_{\rm SM}^\gamma
=  \omega^2_\gamma{\cal  F}_{\rm SM}^\gamma$ for (a)  $e^+d\rightarrow
e^+d+\gamma$  (at  fixed   $\wh{\phi}_\gamma  =  0^{\circ}$)  and  (b)
$e^+u\rightarrow  e^+u+\gamma$  (at fixed  $\wh{\theta}_\gamma  \simeq
78.46^{\circ}$)  for different photon energies  ($\omega_\gamma  = 10,
20,  30,   40$~GeV).  The  outgoing   quark   direction  is  fixed  at
$\Theta_q=90^{\circ}$    with    energy    $E^\prime_q=100$~GeV.   The
directions  of the  incoming  and  outgoing  quarks  and  leptons  are
indicated.}
\label{fig:udxcms}
\end{center}
\end{figure}

In  Fig.~\ref{fig:udxcms}  we show the  dimensionless  quantity ${\cal
N}_{\rm  SM}^\gamma =  \omega_\gamma^2{\cal  F}_{\rm  SM}^\gamma$  for
different  photon energies and fixed  final--state  quark  kinematics.
The figures (a) and (b) correspond  respectively to slices through the
$(\phi_\gamma,\theta_\gamma)$  plane according to the positions of the
soft--photon  Type 1 and 2 radiation  zeros of the previous  sections.
As the photon  energy  increases,  there is a systematic  shift in the
positions  of  the  zeros.  As  radiation  zeros  are  semi--classical
effects due to destructive interference, it is easy to understand that
fixing  the  position  of  the  outgoing   quark  and   simultaneously
increasing $\omega_\gamma$ shifts the interference regions between the
participating  charged particles as the outgoing positron must balance
energy and momentum and thus changes its  relative  orientation.  Thus
the asymmetric  $\omega_\gamma$  dependence of the two radiation zeros
in  Fig.~\ref{fig:udxcms}a  is due simply to our  choice of fixing the
final--state   quark  direction  rather  than  the  direction  of  the
scattered  positron.  The zero in the  quadrant  between  the  (fixed)
incoming   positron  and  outgoing  quark   directions  is  relatively
insensitive  to the  changes  in the  positron  direction  induced  by
varying  $\omega_\gamma$.  The other zero follows the direction of the
outgoing positron as $\omega_\gamma$  increases.  The same effect also
explains the symmetric  dependence of the two radiation  zeros for the
process  $e^+u   \rightarrow   e^+u+\gamma$.  The  zeros  are  located
symmetrically above and below the  scattering plane and are influenced
equally by changes in the scattered positron direction.

\begin{figure}[t]
\begin{center}
\mbox{\epsfig{figure=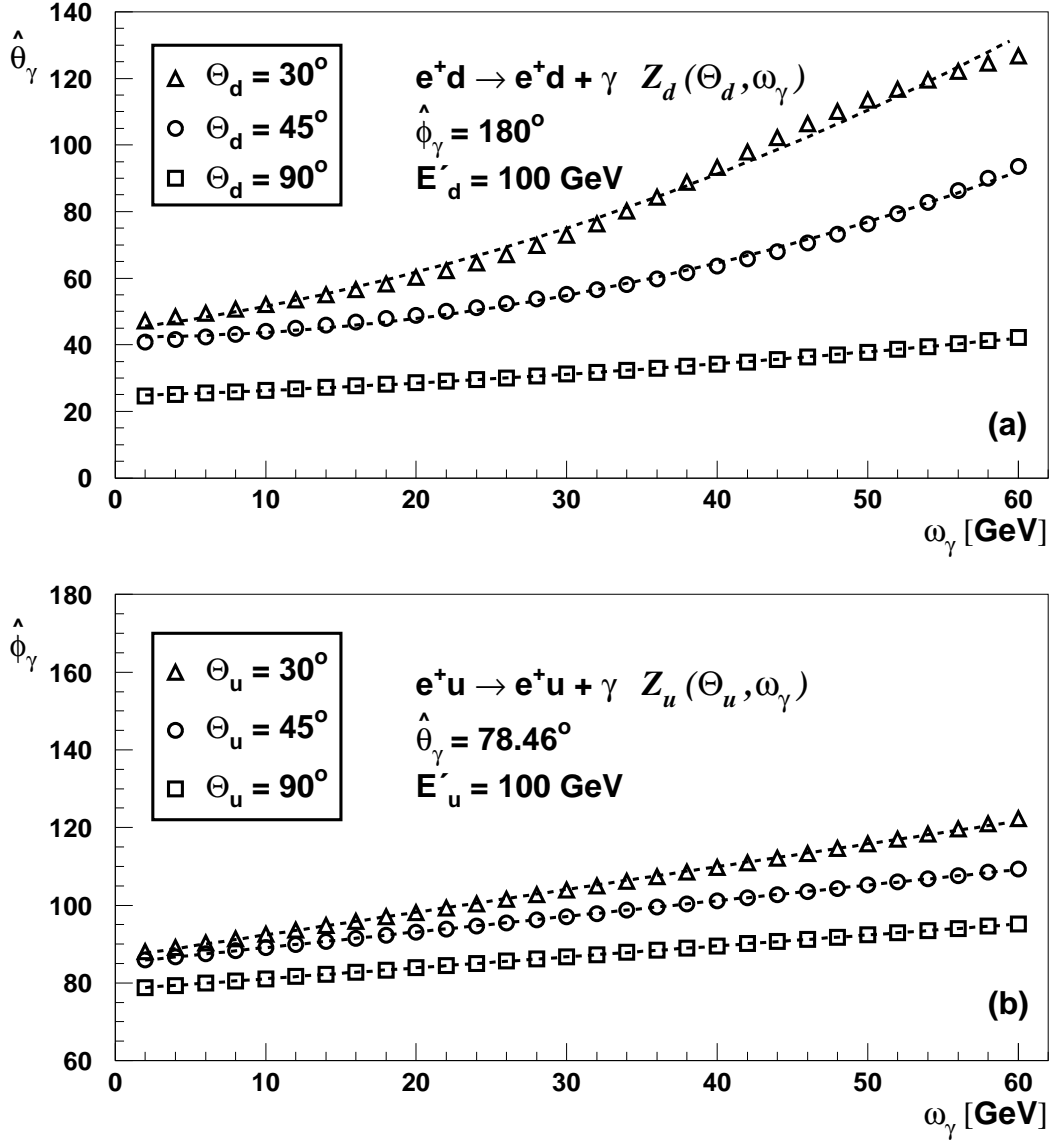,width=15.5cm}}
\caption[]{The  positions  of the  radiation  zeros  ${\cal Z}_q$ as a
function  of the quark  scattering  angle  $\Theta_q$  and the  photon
energy  $\omega_\gamma$  for  (a)  $e^+d$  scattering  and (b)  $e^+u$
scattering.  The   analytic   results  for  the   soft--photon   limit
$(\omega_\gamma\rightarrow       0)$      are       summarised      in
Table~\ref{tab:cmsxzeros}.  The dashed  lines are a polynomial  fit in
the photon energy for given  $\Theta_q$.  Note that in (a) we employ a
second--order  fit whereas in (b) a  first--order  fit is  sufficient.
The fit parameters are listed in the text.}
\label{fig:omegadep}
\end{center}
\end{figure}

Figs.~\ref{fig:omegadep}a,b   show  the  positions\footnote{The  exact
locations of the zeros are  determined by a numerical  procedure.}  of
the  radiation  zeros  ${\cal  Z}_d$  and  ${\cal  Z}_u$  for  the two
processes  as a  function  of  the  photon  energy  at  various  fixed
$\Theta_q$.  ${\cal  Z}_d(\Theta_d,\omega_\gamma)$  is  located in the
quadrant between the outgoing  positron and the incoming $d$ quark and
${\cal   Z}_u(\Theta_u,\omega_\gamma)$  is  located  in  the  quadrant
between  the  outgoing  $u$  quark  and  the  outgoing  positron  (see
Figs.~\ref{fig:udxcms}a,b).  The values on the axes at  $\omega_\gamma
= 0$ coincide  with the  analytic  results  obtained  previously  (see
Table~\ref{tab:cmsxzeros}).

The dashed lines in Figs.~\ref{fig:omegadep}a,b  are simple polynomial
fits.  For $e^+d$  scattering  we fit  $\wh{\theta}_\gamma$  for fixed
$\wh{\phi}_\gamma = 180^{\circ}$ using a quadratic polynomial,
\beq \label{eq:fitd}
{\cal Z}_d(\Theta_d,\omega_\gamma) =  
{\cal Z}_d^0(\Theta_d) 
+ d_1\omega_\gamma 
+ d_2\omega^2_\gamma,
\eeq
where ${\cal Z}_d^0(\Theta_d)$ corresponds to the soft--photon results
listed in  Table~\ref{tab:cmsxzeros}.  The radiation  zeros for $e^+u$
scattering   (i.e.   a   fit   for    $\wh{\phi}_\gamma$    at   fixed
$\wh{\theta}_\gamma   =  78.46^{\circ}$)  can  be  approximated  by  a
first--order polynomial
\beq \label{eq:fitu}
{\cal Z}_u(\Theta_u,\omega_\gamma) =  
{\cal Z}_u^0(\Theta_u) 
+ u_1\omega_\gamma.
\eeq
The results of the fit are presented in Table~\ref{tab:fitxzeros}.
\begin{table}[htb]
\begin{center}
\begin{tabular}{|c||l|l|l||l|l|}                     \hline  
                                                     & 
\multicolumn{3}{c||}{\shift $d$--quarks}             & 
\multicolumn{2}{c|}{$u$--quarks}                  \\ \hline
\shift$\Theta_q$                                     & 
$d_1$                                                & 
$d_2$                                                & 
$\chi^2$                                             & 
$u_1$                                                & 
$\chi^2$                                          \\ \hline 
\shift$30^{\circ}$ & $0.576$ & $0.015$ & $5.91$ 
                   & $0.59$  & $5.3\times 10^{-2}$ \\ \hline
\shift$45^{\circ}$ & $0.012$ & $0.014$ & $0.76$ 
                   & $0.40$  & $3.0\times 10^{-3}$ \\ \hline
\shift$60^{\circ}$ & $0.149$ & $0.002$ & $0.01$ 
                   & $0.28$  & $1.3\times 10^{-4}$ \\ \hline
\end{tabular}
\caption[]{{\em  Fits for the  $\omega_\gamma$  dependence  of the two
selected   radiation   zeros   shown  in   Figs.~\ref{fig:omegadep}a,b
according   to   the   definitions   given   in   Eqs.~(\ref{eq:fitd},
\ref{eq:fitu}).}}
\label{tab:fitxzeros}
\end{center}
\end{table} 

\begin{figure}[t]
\begin{center}
\mbox{\epsfig{figure=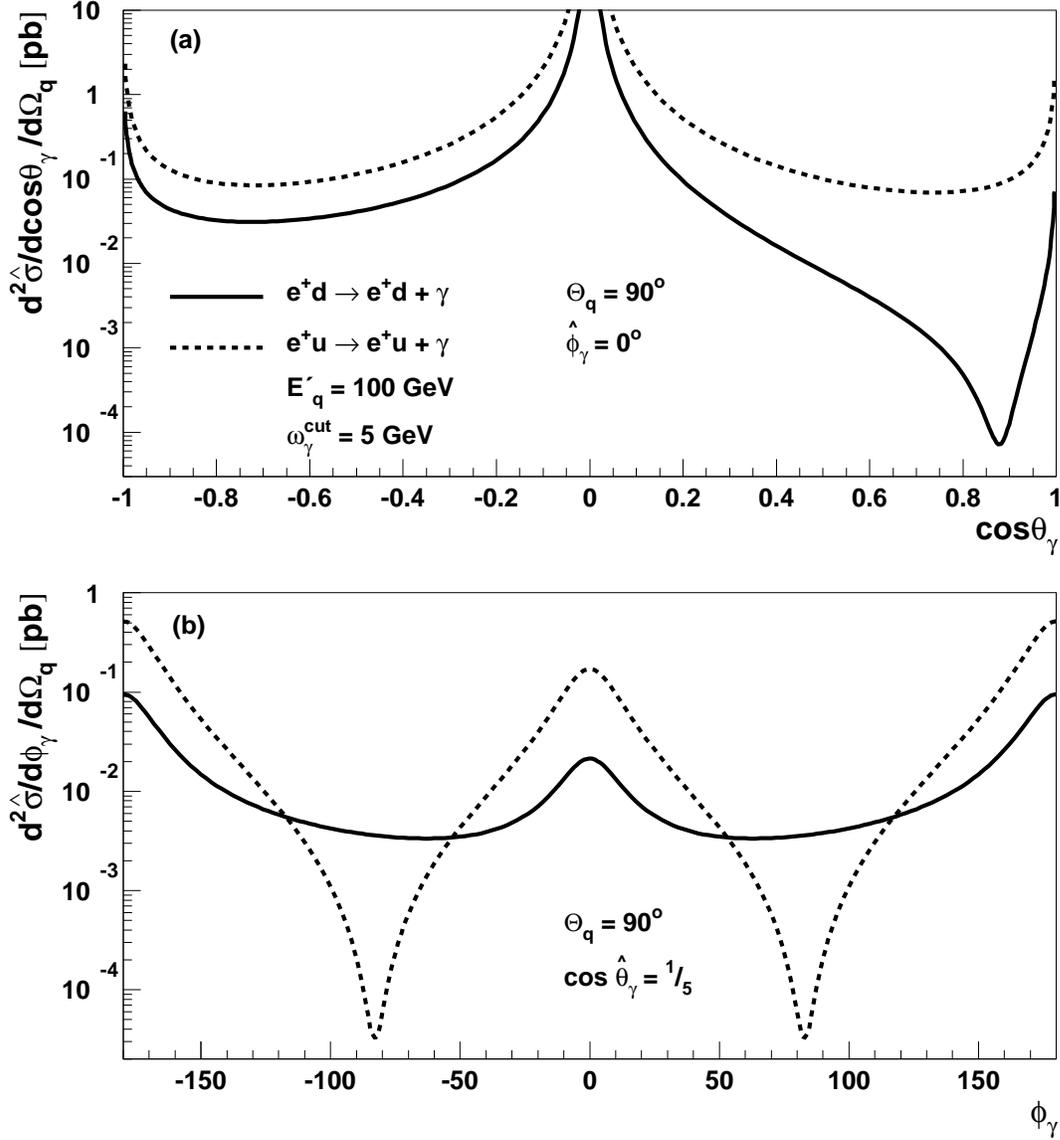,width=15.5cm}}
\caption[]{The subprocess differential cross section $d^2\hat{\sigma}/
d\Omega_\gamma  d\Omega_q$  of  Eq.~(\ref{eq:diffcross})   for  c.m.s.
$e^+d$ scattering (solid lines) and $e^+u$ scattering  (dashed lines).
We again  choose  those  slices  through  the photon  parameter  space
$(\phi_\gamma,\theta_\gamma)$ that contain radiation zeros in the soft
limit  (i.e.  a  choice  of  $\wh{\phi}_\gamma=0^{\circ}$  in (a)  and
$\cos\wh{\theta}_\gamma=1/5$ in (b)).  Note that we integrate over the
photon  energy  $\omega_\gamma$  and fix the  position of the outgoing
quark at $\Theta_q=  90^{\circ}$  with energy  $E^\prime_q = 100$~GeV.
In (a) we impose an additional  angular cut of $5^{\circ}$  around the
beam line.}
\label{fig:cmsxint}
\end{center}
\end{figure}

As  a  final   exercise  in  our  c.m.s.  studies  we  calculate   the
differential cross section for the two subprocesses.  The general form
of the  differential  subprocess  cross  section in the $e^+q$  c.m.s.
frame may be written as
\beq \label{eq:diffcross}
\frac{d^2\hat{\sigma}}{d\Omega_\gamma d\Omega_q}
\left( eq \rightarrow eq+\gamma\right)              = 
\frac{2}{(4\pi)^5}
\int\limits_{\omega_\gamma^{\rm cut}} d\omega_\gamma \;
\frac{E^{\prime 2}_q \omega_\gamma}{\hat{s}^{3/2} 
|\sqrt{\hat{s}}/2 - \omega_\gamma|} \;
\matz(eq\rightarrow eq+\gamma),
\eeq
where
\beq \label{eq:Eqdefn}
E^\prime_q = \frac{ \hat{s} - 2 \sqrt{\hat{s}} \omega_\gamma}
      {2 \sqrt{\hat{s}} - 2 \omega_\gamma(1-\cos\theta_{q\gamma})  }.
\eeq
The integration over $\omega_\gamma$ smears out the radiation zeros to
form a sharp {\em dip} in the cross  section.  Since the cross section
decreases rapidly with increasing $\omega_\gamma$, the dip is close to
the  location  of the zero  corresponding  to fixed  $\omega_\gamma  =
\omega_\gamma^{\rm  cut}$.  The distributions for the two subprocesses
are shown in Figs.~\ref{fig:cmsxint}a,b   for $\omega_\gamma^{\rm cut}
= 5$~GeV.  Note that we have also  imposed an angular  cut around  the
beam line of  $5^{\circ}$ in  Fig.~\ref{fig:cmsxint}a.  The transition
from  radiation  zeros  to  radiation  dips  can be seen by  comparing
Figs.~\ref{fig:omegadep}  and      \ref{fig:cmsxint}.  Choosing larger
values of  $\omega_\gamma^{\rm  cut}$  shifts  the  radiation  dips to
higher values of $\wh{\phi}_\gamma$  and  $\wh{\theta}_\gamma$  at the
same  time  decreasing  the  overall  value  of the  subprocess  cross
section.

\section{Radiation zeros at HERA}

In this section we shall discuss the possible observation of radiation
zeros at HERA.  To do this we modify the previous  calculation  by (a)
moving to the HERA lab frame, (b)  including  the parton  distribution
functions,  and (c) summing over all flavours of quarks in the initial
state.

In neutral current DIS the cleanest way to reconstruct  the kinematics
of a given  event is by  measuring  the  energy  $E^\prime_e$  and the
laboratory angle  $\Theta_e^{\rm  lab}$ of the outgoing  positron.  In
terms of the Bjorken  scaling  variables $x$ and $y$ we may write (see
for example Refs.~\cite{ZEUS95,H1.95})
\begin{eqnarray}
y    &=& 
1 - \frac{E^\prime_e}{2E_e} 
\left( 1 - \cos\Theta^{\rm lab}_e\ \right) , 
\label{eq:yvar} 
\\ 
x    &=& 
\frac{1}{y} \frac{E^\prime_e}{2E_p} 
\left( 1 + \cos\Theta^{\rm lab}_e\ \right) , 
\label{eq:xvar} 
\\ 
Q^2    &=& 
xys , 
\label{eq:qvar}
\end{eqnarray}
where $E_p$  is  the energy of the  incoming  proton and $s = 4E_eE_p$
is the c.m.s.  energy of the  $e^+p$  system.  The polar  angle of the
positron  $\Theta_e^{\rm lab}$ is defined with respect to the incident
proton beam direction.  The precision of the $y$ measurement typically
degrades as $1/y$, and thus one  naturally  assumes  $y\gapprox  0.05$
\cite{H1.95}.

Since we are  interested in DIS events with an additional  hard photon
emitted at different  angles in phase  space, the natural  quantity to
consider is the triple--differential  cross section $d^3\sigma/dy dQ^2
d\Omega_\gamma^{\rm lab}$.  In the HERA lab frame this is given by
\begin{eqnarray}
\nonumber
\label{eq:heralab}
\frac{d^3\sigma}{dydQ^2d\Omega_\gamma^{\rm lab}} ( e + p 
&\rightarrow & 
e+q+\gamma+X) = \frac{1}{256 \pi^4 s}\, \sum_q \\
&\times& 
\int\limits_{\omega_\gamma^{\rm cut}} d\omega_\gamma \;
\frac{\omega_\gamma}{ \xi_q (Q^2/x - 2 p\cdot k)}\; 
\matz(eq\rightarrow eq+\gamma)\; f_{q/p}(\xi_q,Q^2),
\end{eqnarray}
where
\beq
\label{eq:xparton}
\xi_q = \frac{Q^2 - 2 q\cdot k}{Q^2/x - 2 p\cdot k}
\geq x. 
\eeq
\begin{table}[htb]
\begin{center}
\begin{tabular}{|c|c|c|c|} \hline
\shift $y$  & $x$  & $\Theta_e^{\rm lab}$ & $E^\prime_e$ \\ \hline 
\shift 0.20   &  0.55  & $52.4^{\circ}$      & $112.9$~GeV  \\ \hline 
\shift 0.40   &  0.27  & $46.2^{\circ}$      & $107.4$~GeV  \\ \hline 
\shift 0.60   &  0.18  & $38.4^{\circ}$      & $101.9$~GeV  \\ \hline 
\shift 0.80   &  0.14  & $27.6^{\circ}$      & $ 96.4$~GeV  \\ \hline 
\end{tabular}
\caption[]{{\em  Typical values of the scattered  positron  energy and
angle for our parameter choice $Q^2=10^4$~{\rm  GeV}$^2$ and different
values of $y$.}}
\label{tab:xytheta}
\end{center}
\end{table} 
In the  calculations  which follow we choose $E_e = 27.5$~GeV,  $E_p =
820$~GeV  and  neglect  all quark and  lepton  masses.  We again  take
$\omega_\gamma^{\rm  cut} = 5$~GeV  for the lower  limit of the photon
energy.  For the quark distribution functions  $f_{q/p}(\xi_q,Q^2)$ we
use   the    MRS(A$^\prime$)    set   of   partons    introduced    in
Ref.~\cite{MRS95},        with       QCD        scale        parameter
$\Lambda^{N_f=4}_{\overline{\rm   MS}}  =  231$~MeV  corresponding  to
$\as(M^2_Z)  =  0.113$.  In  order  to  stay  in  the   valence--quark
scattering region (i.e.  large $\xi_q$), where we expect the radiation
zeros to be most  visible, we choose  $Q^2 =  10^4$~GeV$^2$  and $y\in
[0.1,1.0]$.  Typical  values  for  $x$  and  the  positron   variables
$\Theta_e^{\rm     lab}$    and    $E^\prime_e$    are    listed    in
Table~\ref{tab:xytheta}.

As we move from the $e^+q$  c.m.s.  frame to the HERA lab  frame,  all
four--momenta are boosted along the beam  direction. Although this has
no effect on the  azimuthal  angles,  the polar  angles  and hence the
locations of all radiation zeros, in particular  $\wh{\theta}_\gamma$,
are  changed.  The  simplest  consequence  of this is that the  $e^+d$
radiation  scattering  zeros remain located in the scattering plane at
$\wh{\phi}_\gamma=0^{\circ}$ and $180^{\circ}$.  To find the locations
of the radiation zeros for process  (\ref{eq:reaced}) we therefore fix
$\wh{\phi}_\gamma   =  0^{\circ}$  and  numerically   determine  their
positions in $\theta_\gamma$.

\subsection{Radiation zeros for $d$ quark scattering}

\begin{figure}[t]
\begin{center}
\mbox{\epsfig{figure=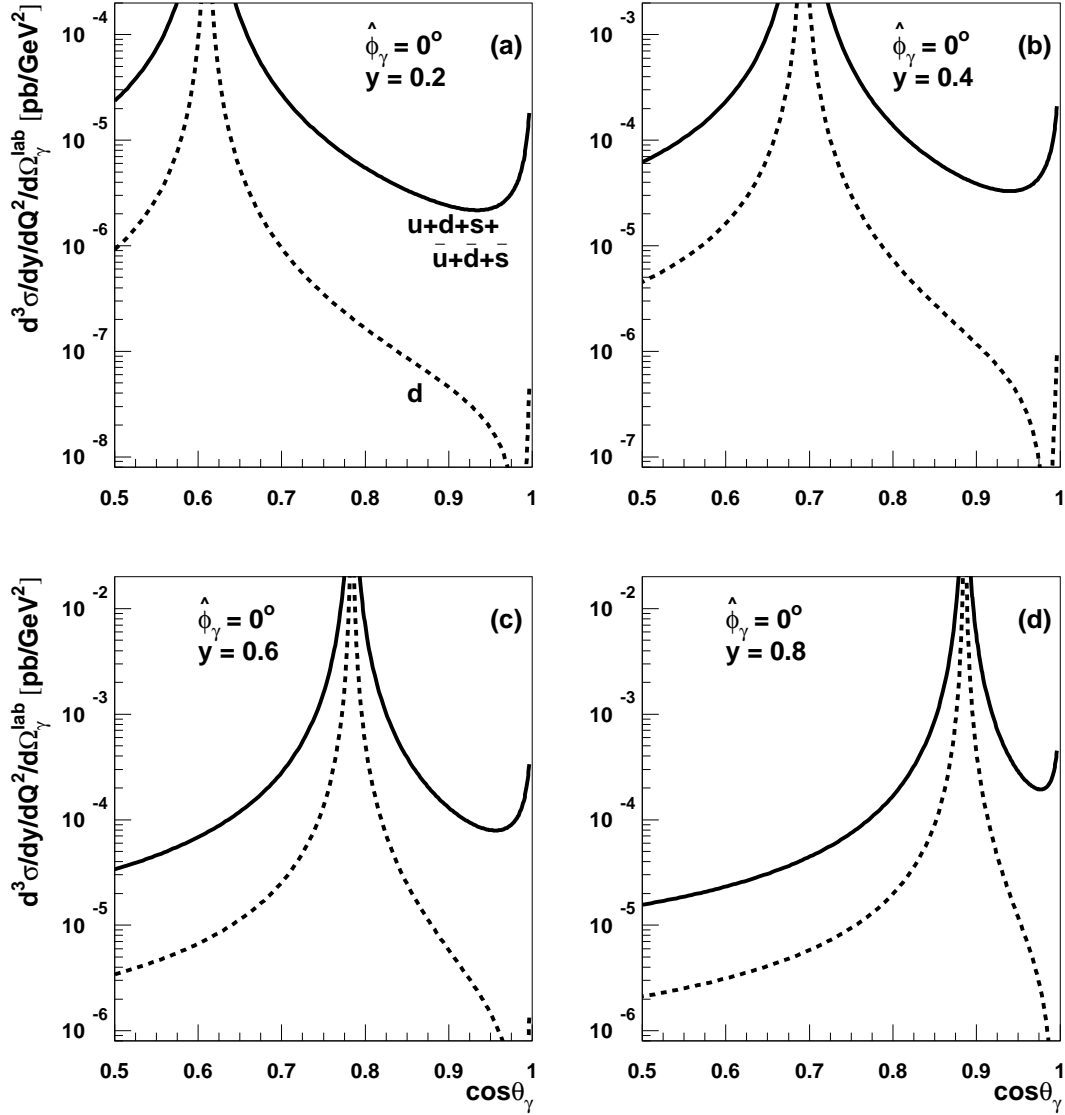,width=15.5cm}}
\caption[]{The  positions of the radiation  dips for the process $e^+d
\rightarrow  e^+d + \gamma$ (dashed lines) for different values of $y$
for $\theta_\gamma \in [5^{\circ},60^{\circ}]$ (hemisphere of outgoing
$e^+$).  A cut of  $5^{\circ}$  around the beam line is  imposed.  The
radiation zeros and thus the radiation dips for this process are again
located  within  the plane  $(\wh{\phi}_\gamma=0^{\circ})$.  The $Q^2$
value  is  $10^4$~GeV$^2$.  The  solid  lines  show  the  sum  of  the
contributions  from $u,d,s$ quarks and antiquarks.  The divergences in
the   plots   show   the   positions   of  the   outgoing   $e^+$   at
$\cos\theta_\gamma  =  \cos\Theta_e^{\rm  lab}$  with the  values  for
$\Theta_e^{\rm lab}$ given in Table~\ref{tab:xytheta}.}
\label{fig:labdzeros}
\end{center}
\end{figure}

In   Fig.~\ref{fig:labdzeros}   we  present   the  cross   section  of
Eq.~(\ref{eq:heralab})  for the process $ e^+ + p \to e^+ + \mbox{jet}
+ \gamma + X$ via $e^+d \rightarrow  e^+d+\gamma$  scattering  (dashed
line) as well as via the sum over all subprocesses  $e^+q  \rightarrow
e^+q +\gamma$ with  $q=u,d,s,\bar{u},\bar{d}$  and $\bar{s}$.  We have
chosen to focus on the  radiation  zero  located  between the incoming
quark and outgoing positron.  We fix $Q^2 = 10^4$~GeV$^2$ and vary $y$
from    $y=0.2$   in    Fig.~\ref{fig:labdzeros}a    to   $y=0.8$   in
Fig.~\ref{fig:labdzeros}d,  which  corresponds  to $x$  values  in the
region  $0.1<x<0.6$  (cf.  Table~\ref{tab:xytheta}).  Again we observe
radiation dips instead of radiation zeros due to the integration  over
the photon energy.  Increasing  $y$ pulls the radiation dips closer to
the beam  line  and  thus  makes  their  observation  more  difficult.
Already    at    $y=0.2$     the    $e^+d$     radiation     dip    in
Fig.~\ref{fig:labdzeros}a      is      only      about      $14^\circ$
($\cos\wh{\theta}_\gamma  \simeq  0.97$)  from the beam line, and gets
even  closer  with  increasing  $y$.  Note  that  we  impose  a cut of
$5^{\circ}$ around the beam line.  Increasing $y$ means decreasing the
polar  angle  of  the  outgoing  positron  $\Theta_e^{\rm  lab}$  (cf.
Table~\ref{tab:xytheta}).  Thus the zone of  destructive  interference
approaches the beam line as the $e^+$  approaches  the beam line.  The
conclusion is that  observation  of the  radiation  dips in the sector
between the incoming quark and outgoing $e^+$ in high--$Q^2$ events is
only possible for small values of $y$.

\begin{figure}[t]
\begin{center}
\mbox{\epsfig{figure=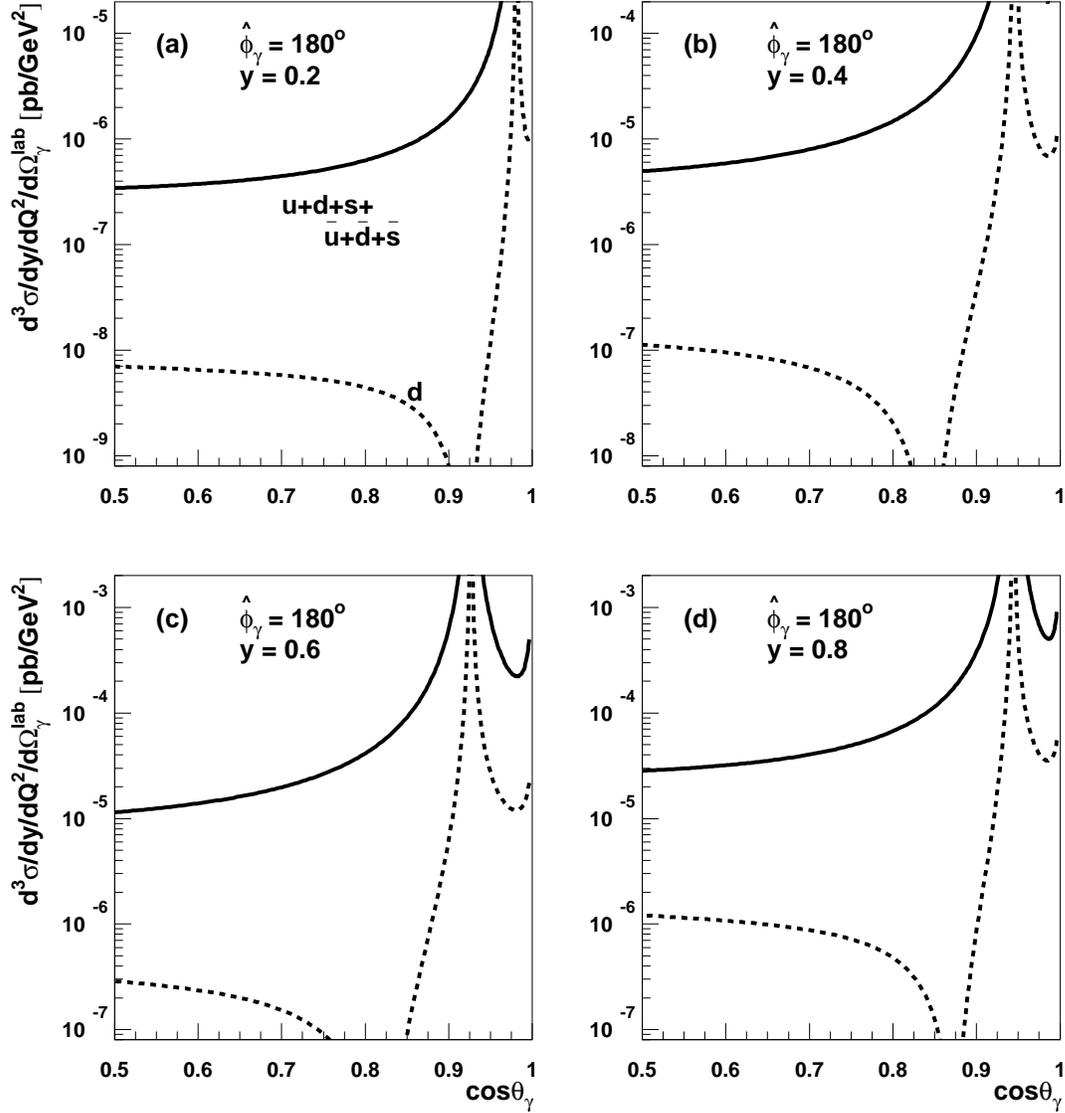,width=15.5cm}}
\caption[]{The  positions of the radiation  dips for the process $e^+d
\rightarrow  e^+d + \gamma$ (dashed lines) for different values of $y$
for $\theta_\gamma \in [5^{\circ},60^{\circ}]$  (the hemisphere of the
outgoing  quark).  A cut  of  $5^{\circ}$  around  the  beam  line  is
imposed.  Note that $\wh{\phi}_\gamma=180^{\circ}$.}
\label{fig:labdzeros2}
\end{center}
\end{figure}

The second  radiation zero we found in our studies was located between
the    incoming     positron    and    the    outgoing    quark.    In
Fig.~\ref{fig:labdzeros2}  we display this region again for  processes
only  involving  $d$ quarks  (dashed  lines) as well as for  processes
involving  all  light  quark  and  antiquark   flavours.  The  obvious
singularities   in   Figs.~\ref{fig:labdzeros2}a--d   are   caused  by
collinearity  of the photon with the outgoing  quark.  Now the problem
is that the  zeros  are  close  (always  within  $10^{\circ}$)  to the
outgoing  quark  jet, even  though  the  radiation  dips here are well
separated from the beam line ($\simeq 35^{\circ}$ for $y=0.6$).

A  more  serious  problem  evident  in  Figs.~\ref{fig:labdzeros}  and
\ref{fig:labdzeros2}  is the enormous  background from the other quark
scattering  subprocesses, which completely fills in the radiation dip.
We   observe  a  ratio   (away  from  the   singularities)   of  ${\rm
signal/background}\;\simeq  1/(200-300)$.  The  dominance  of the  $u$
quark  contribution is striking.  For the given values of $x$ and thus
$\xi_q$ (cf.  Eq.~(\ref{eq:xparton})) we find the following ratios for
the MRS(A$^\prime$) parton distributions at $Q^2 =10^4$~GeV$^2$:
\begin{eqnarray}
\xi_q = 0.1: \hspace{0.8cm} &\rightarrow& 
u(\xi_q):d(\xi_q):\bar{d}(\xi_q):\bar{u}(\xi_q)
\simeq 100:60:22:15 , \label{eq:ratio0.1}
\\
\xi_q = 0.6: \hspace{0.8cm} &\rightarrow& 
u(\xi_q):d(\xi_q):\bar{d}(\xi_q):\bar{u}(\xi_q)
\simeq 100:17:1:1 . \label{eq:ratio0.6}
\end{eqnarray}
In addition to these parton  distribution  factors there are the usual
quark charge squared ($e_q^2$)  factors from the leading order $eq \to
eq$  scattering,  which further  enhance the $u$--quark  contribution.
Note  that  the  $s$--quark  contribution  plays a minor  role;  it is
roughly  70\%  of  the  $\bar{u}$   contribution  at  $\xi_q=0.1$  and
comparable  to the latter at higher  values of  $\xi_q$.  Even  though
$d,s$ and $\bar{u}$  quarks all yield radiation dips in the scattering
plane (the  $d$-- and  $s$--quark  zeros  coincide)  none of these are
likely  to be  observable.  The  only  possibility  might be to try to
flavour--tag  the $d$ or $s$ quark jets, for example by selecting only
those jets with a leading negatively charged track.

\subsection{Radiation zeros for $u$ quark scattering}

According  to  the  parton  distribution  hierarchy  presented  in the
previous  section we might  expect  that the Type 1  radiation  zeros,
which we identified with the {\em traditional} radiation zeros already
discussed in the literature, are the most promising for detection.  We
recall that in the  soft--photon  limit and in the c.m.s.  frame these
zeros are located at fixed  polar  angle  $\cos\wh{\theta}_\gamma=1/5$
(cf.  Eq.~(\ref{eq:uzeros})).  Their  position  in  $\phi_\gamma$  may
then be directly  computed  using  Eq.~(\ref{eq:uzerosphi}).  We found
that they are located well outside the  scattering  plane  (except for
$\Theta_q = 2  \wh{\theta}_\gamma  =  2\cos^{-1}(1/5))$  as  discussed
earlier.  Integrating over the photon energy $\omega_\gamma$ and using
exact  $2\rightarrow 3$ kinematics slightly shifts the position of the
corresponding  radiation  dips.  The  $\omega_\gamma$  dependence  for
different      kinematical      situations      was      shown      in
Fig.~\ref{fig:omegadep}b.

\begin{figure}[t]
\begin{center}
\mbox{\epsfig{figure=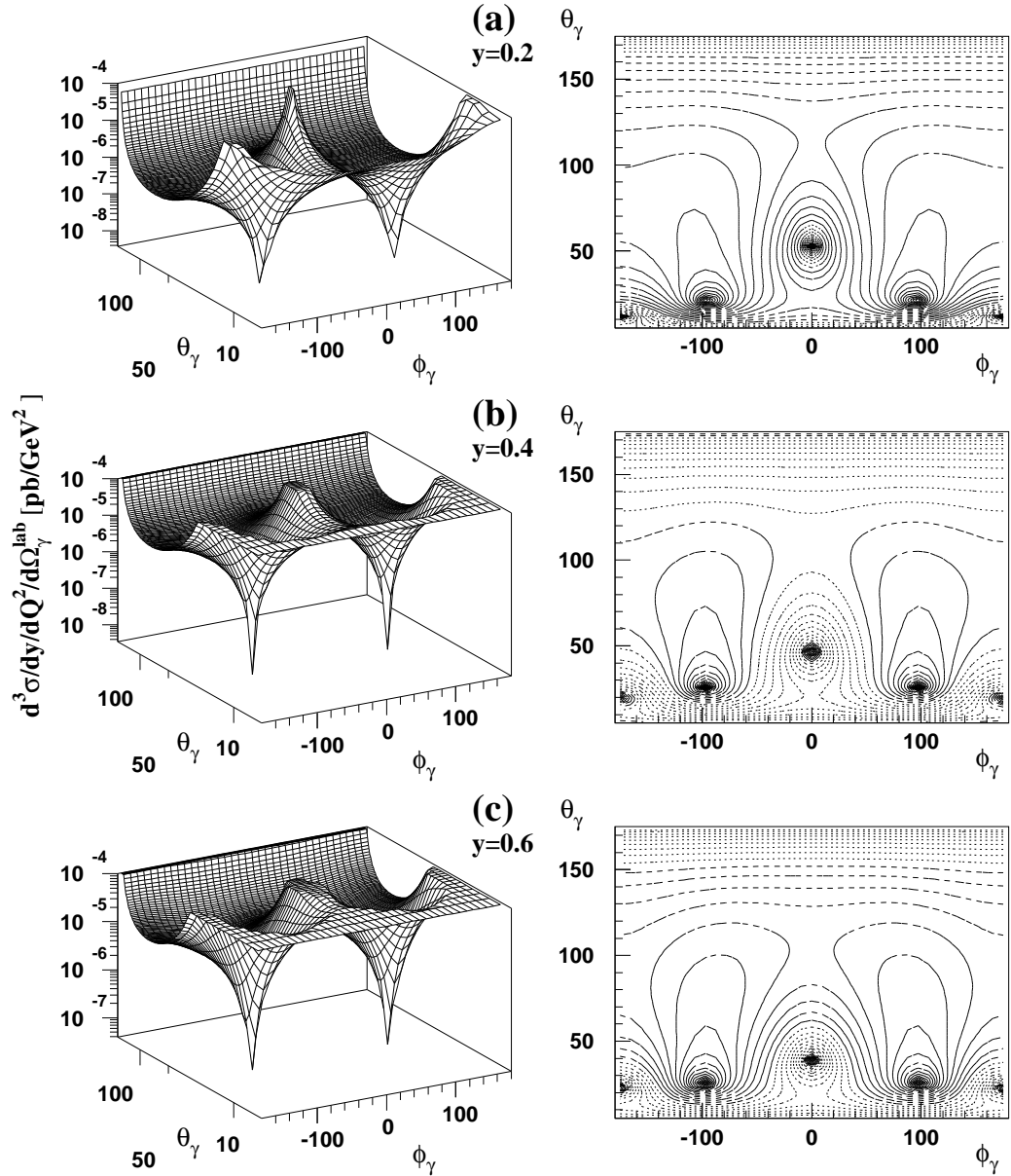,width=15cm}}
\caption[]{The  differential  cross section of  Eq.~(\ref{eq:heralab})
for the process $e^+u \rightarrow e^+u + \gamma$ in the $(\phi_\gamma,
\theta_\gamma)$  phase space of the emitted photon.  The corresponding
contour  plots are shown on the  right--hand  side.  For fixed  $Q^2 =
10^4$~GeV$^2$  we  vary  $y$  (defined  in  Eq.~(\ref{eq:yvar}))  from
$y=0.2$ in (a) and  $y=0.4$ in (b) to $y=0.6$  in (c).  The  kinematic
variables $x, E_e^\prime$ and $\Theta_e^{\rm  lab}$ for each $y$ value
can be read off from  Table~\ref{tab:xytheta}.  Note that we introduce
in the surface  plots on the  left--hand  side a logarithmic  scale in
$\theta_\gamma$.  The radiation  dips are symmetric in  $\phi_\gamma$.
Again we impose a  $5^{\circ}$  cut  around  the beam  line,  and thus
$\theta_\gamma\in [5^{\circ},175^{\circ}]$.}
\label{fig:ufull0}
\end{center}
\end{figure}

Moving to the HERA lab frame  boosts the polar  angles and changes the
position of the radiation dips for $e^+u  \rightarrow  e^+u + \gamma$.
In  Figs.~\ref{fig:ufull0}a--c  we show the differential cross section
of   Eq.~(\ref{eq:heralab})   for   this   process   over   the   full
$(\phi_\gamma,\theta_\gamma)$     space.    As    before     we    fix
$Q^2=10^4$~GeV$^2$  and chose the three $y$ values:  0.2, 0.4 and 0.6.
We impose  cuts of  $5^{\circ}$  around  the beam line (by  definition
located at $\theta_\gamma = 0^{\circ}$ and $180^{\circ}$)  and cut the
differential cross section at $ d\sigma < 10^{-4}$~pb/GeV$^2$ to avoid
the collinear singularities along the directions of the outgoing $e^+$
(located at  $\phi_\gamma=0^{\circ}$)  and the outgoing  $u$ quark (at
$\phi_\gamma=\pm  180^{\circ}$).  We see  that  the  positions  of the
zeros are still  symmetric in  $\phi_\gamma$,  as expected.  Note that
since  the  collinear  singularities  and the  radiation  dips tend to
concentrate around small values of $\theta_\gamma$, we have introduced
a  logarithmic  scale for  $\theta_\gamma$  in the  three--dimensional
plots of Fig.~\ref{fig:ufull0}.

We can  numerically  locate the positions of the radiation dips in the
$(\phi_\gamma,\theta_\gamma)$  phase  space for our different  choices
of $y$:

\parbox{2.0cm}{\begin{eqnarray*} 
\hspace{3.0cm}                   y = 0.2 \; &\rightarrow& \\
\hspace{3.0cm}                   y = 0.4 \; &\rightarrow& \\ 
\hspace{3.0cm}                   y = 0.6 \; &\rightarrow& 
                                 \end{eqnarray*}} \hfill
\parbox{4cm}{
\begin{eqnarray*}
                \wh{\phi}_\gamma &\simeq& \pm 97.2^{\circ} \, ,
                 \\
                 \wh{\phi}_\gamma &\simeq& \pm 100.4^{\circ} \, , 
                  \\
                   \label{eq:y0.6}
                 \wh{\phi}_\gamma &\simeq& \pm 102.5^{\circ} , 
                  \end{eqnarray*}
}
\parbox{6cm}{
\begin{eqnarray} \nonumber
\wh{\theta}_\gamma &\simeq& 20.6^{\circ}; \hspace{3cm}
\\ \nonumber
\wh{\theta}_\gamma &\simeq& 24.9^{\circ}; \hspace{3cm}
\\ 
\wh{\theta}_\gamma &\simeq& 25.2^{\circ}. \hspace{3cm}
\end{eqnarray}
}

It is straightforward  to verify that the radiation dips, if projected
onto the  scattering  plane,  lie within  the  quadrants  between  the
incoming  (outgoing) $e^+$ and the outgoing (incoming) quark, the zone
of  destructive  interference.  As  Fig.~\ref{fig:ufull0}  shows,  the
radiation  dips are clustered  quite close to the (beam)  direction of
the incoming quark  ($\theta_\gamma=  0^\circ$)  which is particularly
true for high--$Q^2$  events  (back--scattered  positron).  As we have
already   pointed   out,  they  are  also  within   $10^{\circ}$   (in
$\theta_\gamma$)  of the  final--state  quark jet.  However, they {\em
are}   well--separated   from   the   outgoing   particles   when  the
$\phi_\gamma$  angle is taken into account.  It will be very important
to perform  realistic  simulations of these photon  radiation  events,
including jet fragmentation  and detector  effects, to see whether the
dips are indeed observable in practice.

\begin{figure}[t]
\begin{center}
\mbox{\epsfig{figure=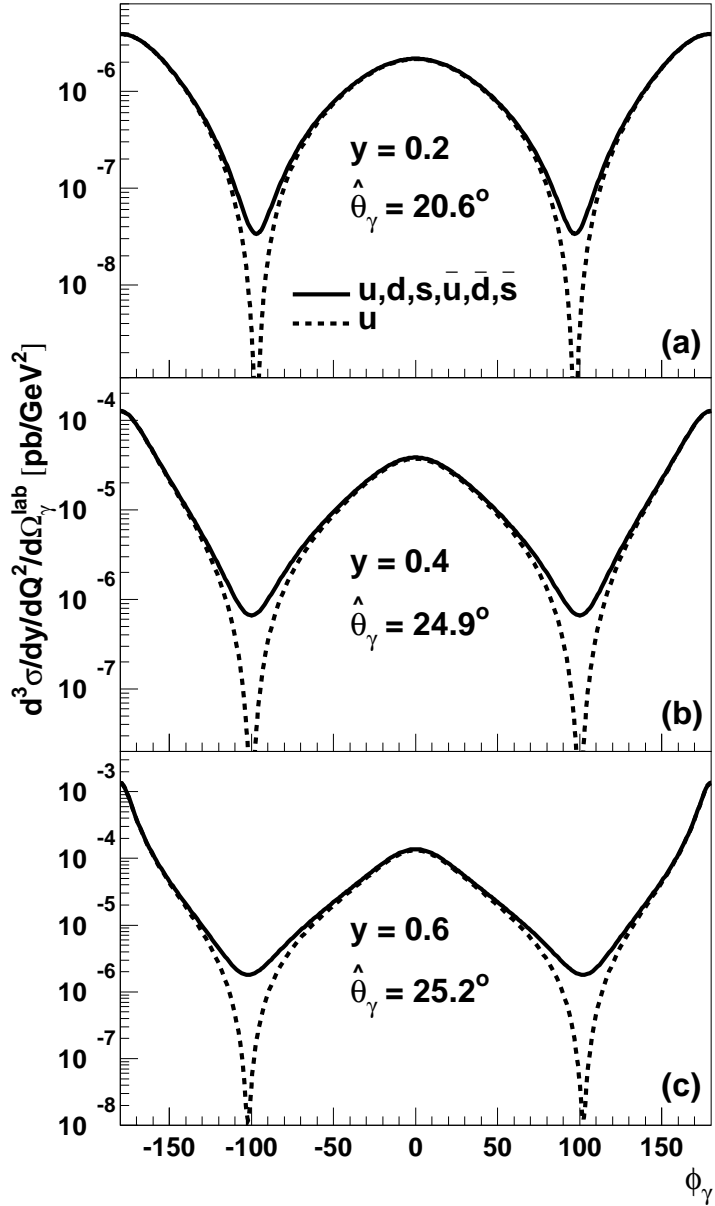,width=11.0cm}}
\caption[]{The  differential  cross section of  Eq.~(\ref{eq:heralab})
for three  different $y$ values at the position of the radiation  dips
$\wh{\theta}_\gamma$  shown in Fig.~\ref{fig:ufull0}  as a function of
the  azimuthal   angle   $\phi_\gamma$.  We  show  the  process  $e^+u
\rightarrow e^+u + \gamma$ (dashed lines) as well as the  contribution
(solid  lines)  of  all  light  quark  flavours  ($u,d,s$  quarks  and
antiquarks).}
\label{fig:uslice}
\end{center}
\end{figure}

Finally, in Fig.~\ref{fig:uslice} we show the $\phi_\gamma$ dependence
for    slices    through    the    $\wh{\theta}_\gamma$    values   of
Eq.~(\ref{eq:y0.6})   which  define  the  numerical  location  of  the
radiation dips of Fig.~\ref{fig:ufull0}.  We show the contributions of
$u$ quarks only, as well as the contributions  from all light flavours
(i.e.  $u,d$ and $s$ quarks and  antiquarks).  At the critical  values
of $\wh{\phi}_\gamma$ (again given in Eq.~(\ref{eq:y0.6})) the obvious
dips for pure  $u$--quark  scattering  are  somewhat  filled in by the
other `background' (mainly $d$--quark) processes --- the cross section
at the bottom of the dip is increased by about two orders of magnitude
--- although they are still significant.

\subsection{Radiation zeros and `parton shower' models}

\begin{figure}[t]
\begin{center}
\mbox{\epsfig{figure=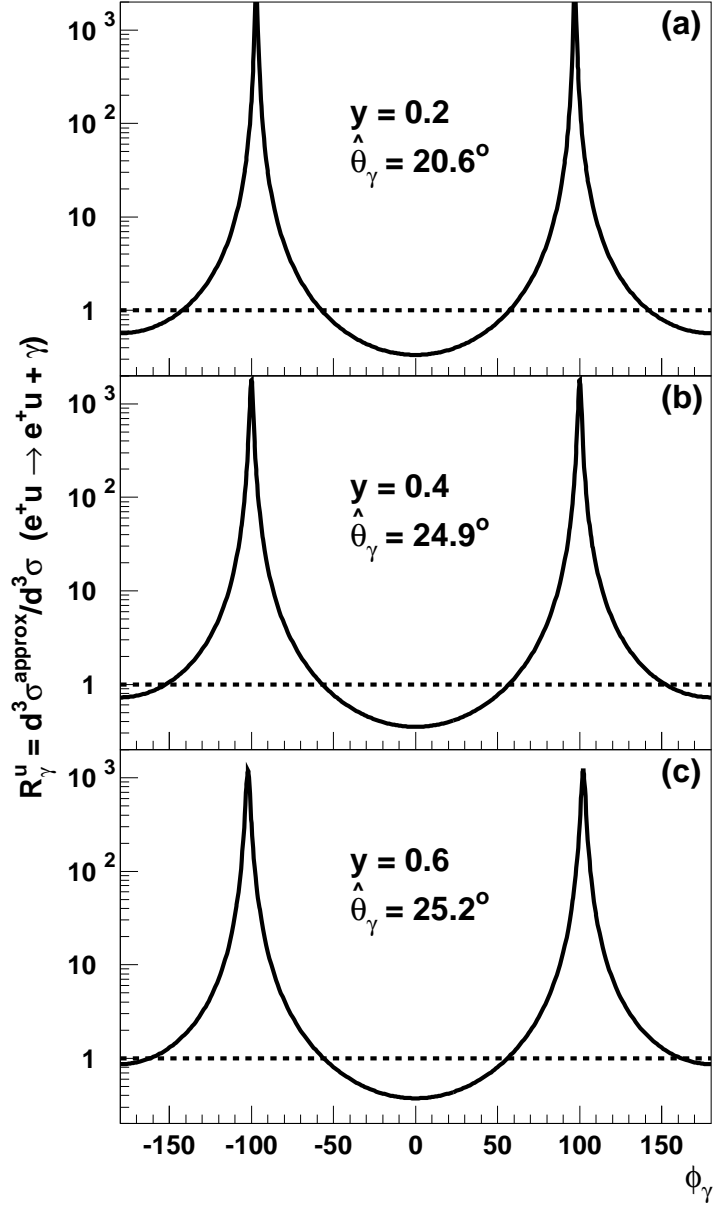,width=11.0cm}}
\caption[]{Same  as  Fig.~\ref{fig:uslice},  but now  for the  process
$e^+u  \rightarrow  e^+u + \gamma$ only.  $R^u_\gamma$ is the ratio of
the differential cross sections of  Eq.~(\ref{eq:heralab}) without and
with interference terms, see Eq.~(\ref{eq:ratio}).}
\label{fig:uratio}
\end{center}
\end{figure}

To  gauge  the  quantitative   significance  of  the  radiation  zeros
described in the previous  sections, and in  particular  to factor out
the effects of phase space  constraints  on the  distributions,  it is
useful to make  comparison  with an  approximate  calculation in which
radiation  zeros are absent.  Parton shower Monte Carlo programs, such
as  {\sc  Herwig}  \cite{Mar92} or  {\sc  Pythia}  \cite{Sjo94},  
are based on the  principle of the
leading--pole  (collinear)  approximation.  In particular  they do not
usually  include  the  interference  effects  which  are  crucial  for
producing radiation zeros in the scattering amplitudes.  We can easily
emulate  such   models by  removing  the  interference  terms from the
antenna  pattern  in  Eq.~(\ref{eq:fsm})  (i.e.  the  terms  linear in
$e_q$):
\beq \label{eq:fsmapprox}
\frac{1}{2}{\cal F}_{\rm SM}^{\gamma \; {\rm approx}} = 
e_q^2[24] + [13],
\eeq
The  approximate  matrix  element  thus  obtained  still  contains the
correct  leading  collinear  singularities  when the photon is emitted
parallel  to  the  incoming  and  outgoing   quarks  and  leptons.  In
Fig.~\ref{fig:uratio} we present the ratio
\beq
\label{eq:ratio}
R^u_\gamma = \frac{d^3\sigma^{\rm approx}}{d^3\sigma} 
\left( e^+u \rightarrow e^+u + \gamma \right),
\eeq
where $d^3\sigma^{\rm approx}/dydQ^2d\Omega^{\rm lab}_\gamma$ includes
the  antenna  pattern  without   interference  terms,  as  defined  in
Eq.~(\ref{eq:fsmapprox}).  Again we slice through $\phi_\gamma$ at the
values    $\wh{\theta}_\gamma$   of   Eq.~(\ref{eq:y0.6})   where   we
numerically  located the positions of the radiation  dips for each $y$
value.  Note that away from the dips the ratio is  ${\cal  O}(1)$,  as
expected.  However  Fig.~\ref{fig:uratio} also shows that close to the
dips the approximate  cross section is up to three orders of magnitude
larger than the exact result, for all $y$ values. In these  particular
regions of phase space, therefore, such `parton--shower'  models would
dramatically overestimate the photon emission cross section.

\section{Conclusions}

The scattering amplitude for the process $eq \to q e +\gamma$ vanishes
for certain configurations of the final--state momenta.  In this paper
we  have  studied  these  radiation  zeros  and  in  particular  their
observability  at HERA.  In addition to the well-known  class of (Type
1)  same--charge  zeros, which have been  discussed in the  pioneering
work of  Refs.~\cite{Bil85}  and  \cite{Don91},  we have  discovered a
second class of (Type 2) zeros located in the $eq$  scattering  plane.
We have so far been unable to find a theorem which leads to conditions
for the existence of such zeros in more general scattering  processes.

Experimentally, one might hope to be able to measure the four--momenta
of the  final--state  lepton,  quark  (jet)  and  photon  sufficiently
accurately that the kinematic configurations which lead to zeros could
be  reconstructed.  However a more realistic  approach,  which we have
adopted  here, is to study DIS $+$  photon  events  for  fixed  lepton
variables  $y$ and $Q^2$ and for a range of  photon  energies  above a
given threshold.  This leads to sharp radiation dips instead of zeros.
We  performed  such a study  using the HERA lab  frame.  Although  the
radiation  dips,  i.e.  the  photon  directions  for  which the  cross
section has a minimum, of both types are quite well separated from the
beam direction and from the  final--state  jet, the $e^+d$  scattering
dips are completely  swamped by the contributions from the other quark
scattering processes.  The $e^+u$ (Type 1) dips offer a more promising
hope of detection, since $e^+u$ scattering is the dominant  subprocess
at high $x$.  With sufficient statistics, we would expect such dips in
the cross section to be observable at HERA.

\vspace{1.0cm}


\noindent {\bf Acknowledgements} \\

We thank  Bernd  L\"ohr  from  the  ZEUS  Collaboration  for  fruitful
discussions   on   various   experimental   aspects.   MH   gratefully
acknowledges     financial     support    in    the    form    of    a
DAAD--Dok\-tor\-an\-den\-sti\-pen\-di\-um  (HSP III).  WJS is grateful
to the Fermilab Theory Group for hospitality  during the completion of
this work.

\newpage



\newpage


\end{document}